\documentclass[12pt,preprint]{aastex}
\usepackage{lscape}

\newcommand{\kms}{km\ s$^{-1}$}

\newcommand{\Msun}{$M_{\sun}$}

\newcommand{\bm}{beam$^{-1}$}
\newcommand{\yr}{yr$^{-1}$}

\newcommand{\ppcsq}{pc$^{-2}$}
\newcommand{\fcon}{f$_{con}$}
\newcommand{\snuc}{$\Sigma_{nuc}$}
\newcommand{\stot}{$\Sigma_{tot}$}
\newcommand{\scrit}{$\Sigma_{crit}$}
\newcommand{\mnuc}{M$_{nuc}$}
\newcommand{\mtot}{M$_{tot}$}

\slugcomment{Submitted to ApJ}

\shorttitle{Bar-Driven Secular Evolution}
\shortauthors{Sheth et al.}

\begin{document}

\title{Secular Evolution Via Bar-Driven Gas Inflow: Results from BIMA SONG}

\author{Kartik Sheth\altaffilmark{1,2,3,4},Stuart
N. Vogel\altaffilmark{2},Michael W. Regan\altaffilmark{3,5},Michele
D. Thornley\altaffilmark{6},Peter J. Teuben\altaffilmark{2}}

\altaffiltext{1}{Spitzer Science Center, MS 220-6, 
California Institute of Technology, Pasadena, CA 91125}

\altaffiltext{2}{Department of Astronomy, University of Maryland,
College Park, MD 20742-2421}

\altaffiltext{3}{Visiting Astronomer, Kitt Peak National Observatory.
KPNO is operated by AURA, Inc.\ under contract to the National Science
Foundation.}

\altaffiltext{4}{Email: kartik@astro.caltech.edu}

\altaffiltext{5}{Space Telescope Science Institute, 3700 San Martin
Drive, Baltimore, MD 21218}

\altaffiltext{6}{Department of Physics, Bucknell University, Lewisburg,
PA 17837}

\begin{abstract}

We present an analysis of the molecular gas distributions in the 29
barred and 15 unbarred spirals in the BIMA CO (J=1--0) Survey of
Nearby Galaxies (SONG).  For galaxies that are bright in CO, we
confirm the conclusion by \citet{sakamoto99b} that barred spirals have
higher molecular gas concentrations in the central kiloparsec.  The
SONG sample also includes 27 galaxies below the CO brightness limit
used by Sakamoto et al.  Even in these less CO-bright galaxies we show
that high central gas concentrations are more common in barred
galaxies, consistent with radial inflow driven by the bar.  However,
there is a significant population of early-type (Sa--Sbc) barred
spirals (6 of 19) that have no molecular gas detected in the nuclear
region, and have very little out to the bar co-rotation radius.  This
suggests that in barred galaxies with gas-deficient nuclear regions
the bar has already driven most of the gas within the bar co-rotation
radius to the nuclear region, where it has been consumed by star
formation.  The median mass of nuclear molecular gas is over four times  
higher in early type bars than in late type (Sc--Sdm) bars.  
Since previous work has shown that the gas consumption rate is an
order of magnitude higher in early type bars, this implies that the
early types have significantly higher bar-driven inflows.  The lower
accretion rates in late type bars can probably be attributed to the
known differences in bar structure between early and late types.
Despite the evidence for bar-driven inflows in both early and late
Hubble type spirals, the data indicate that it is highly unlikely for
a late type galaxy to evolve into an early type via bar-induced gas
inflow.  Nonetheless, secular evolutionary processes are undoubtedly
present, and pseudo-bulges are inevitable; evidence for pseudo-bulges
is likely to be clearest in early-type galaxies because of their high
gas inflow rates and higher star formation activity.

\end{abstract}
\section{Introduction}\label{intro}

Nearly three-quarters of all nearby disk galaxies have a bar
\citep{eskridge00,menendez04}; recent studies suggest that the bar
fraction remains high at z $>$ 0.7 \citep{sheth03}, and is perhaps
constant to a redshift of one \citep{elmegreen04, jogee04b}.  Models
show that the non-axisymmetric bar potential induces large scale
streaming motions in the stars and gas (\citealt{sellwood93} and
references therein, \citealt{a92a,a92b}).  This mixing is responsible
for the shallower chemical abundance gradients observed in barred
spirals compared to unbarred spirals
\citep{vilacostas92,zaritsky94,martin94,martinet97}.  Studies of gas
kinematics in the bar indicate that molecular gas is flowing inwards
down the bar dust lanes \citep{downes96,regan99a,sheth00,sheth02}.
Enhancement of the star formation activity in the nuclei of barred
spirals (e.g., \citealt{sersic67,hawarden86,ho97b}) suggests that
either the conditions at the centers of barred spirals are more
favorable for star formation or these regions have a higher gas
content.  We note that the enhancement in star formation activity is
primarily found in early Hubble type barred spirals \citep{ho97b}.  Ho
et al. divide the galaxies with the Hubble types Sa-Sbc as early
types, and galaxies with type Sc--Sdm as late types; we use this
division throughout the remainder of this paper (see also \S
\ref{data}).  The observational results noted above provide
indirect observational evidence that bars transport gas towards their
centers.

Models also suggest that further inflow of gas by a
``bars-within-bars'' scenario may fuel active galactic nuclei
\citep{shlosman89}, although the evidence for such fueling is weak
\citep{regan99b,knapen00,laine02,lauri04}.  Other studies have
suggested that bar-induced gas inflows may lead to the formation of
new (pseudo-) bulges (e.g., \citealt{kormendy93}, see also recent
review by \citealt{kormendy04}), and evolve late Hubble type spirals
into early Hubble types (e.g., \citealt{friedli95,kenney96,zhang99}).
The most dramatic prediction is that bars may destroy
themselves if they accumulate sufficient mass at their centers
(\citealt{pfen90,friedli93,norman96,bournaud02,athanassoula03,shen04}; see
also review by \citealt{kormendy04} and references therein).

Starburst activity, the formation of a pseudo-bulge, secular evolution
of a galaxy along the Hubble sequence, and the destruction of the bar
are all expected to occur when gas accretes in the nuclear region of a
galaxy.  Is there any direct evidence for excess gas at the center of
barred spirals?  While studies of individual galaxies have shown that
at least some bars have a substantial amount of molecular gas near
their centers (e.g., \citealt{kenney96,turner96}), only one study has
addressed the issue of whether the central regions of barred spirals
are {\it systematically} more gas rich than those in unbarred spirals.
With a sample of twenty (ten barred and ten unbarred) galaxies, the
Owens Valley - Nobeyama Radio Observatory (NRO-OVRO) survey
\citep{sakamoto99a} found that the central concentration (\fcon),
defined as the ratio of the nuclear molecular gas surface density
(\snuc) to the total (disk-averaged) molecular gas surface density
(\stot), is significantly higher in barred spirals than unbarred
spirals \citep{sakamoto99b}.  While \fcon\ increases when gas is
transported to the center, it decreases when gas is consumed by
circumnuclear star formation.  Since barred spirals have elevated
circumnuclear star formation activity (e.g.,
\citealt{sersic67,hawarden86,ho97b}), a priori, one expects {\sl
lower} \fcon\ values in barred spirals, {\sl unless there is a fresh
influx of molecular gas}.  Thus a higher \fcon\ value in barred
spirals than unbarred spirals is strong evidence for bar-induced gas
inflow and accumulation.  Moreover, \fcon\ is robust against biases in galaxy size,
mass, and distance. In addition, given the shallower metallicity
gradients typically found in barred spirals, variations in CO brightness with
metallicity would cause \fcon\ to be {\sl smaller} in barred
than unbarred spirals, the opposite of what is observed \citep{ho97b}.

Although the NRO-OVRO result is statistically significant, the sample
size is modest (20 galaxies) and consists of CO-bright
galaxies\footnote{ The NRO-OVRO survey chose galaxies with integrated
  intensities of I$_{CO}>$ 10 K \kms\ in any one of the 50$\arcsec$
  FCRAO pointings (typically the central pointing). We use the 10 K
  \kms\ selection as the division between CO-bright and CO-faint
  galaxies for the remainder of this paper.}.  The NRO-OVRO sample has
  only four galaxies of Hubble type Sc or later.  Hence it is unclear
  whether the observed difference in \fcon\ between barred and
  unbarred spirals is representative of all spiral galaxies.  Does
  this result extend to CO-faint spirals?  How does it vary with
  Hubble type?  Is there any evidence for secular evolution from late
  to early Hubble type?

With the larger and more diverse sample of spiral galaxies in SONG we
can address these outstanding questions. SONG contains three times as
many bars as the NRO-OVRO study, and 50\% more unbarred spirals.
There are 27 galaxies in SONG with lower CO-brightness than the
NRO-OVRO sample because SONG galaxies were chosen based on an apparent
optical magnitude instead of CO-brightness.  Moreover, SONG has 15
spirals of Hubble type Sc through Sd; thus we can directly evaluate
the influence of bars on late Hubble type spirals.  One other
advantage of SONG is that it imaged the CO emission over the entire
bar region in all but one of the barred spirals; hence we can compare
the gas distribution in the central region to the available gas in the
bar and place the observations in the context of secular evolution.

\section{OBSERVATIONS, DATA REDUCTION \& ANALYSIS}\label{data}

In SONG, the molecular gas distributions (using the CO J=1--0 line)
in 44 nearby spirals (Sa-Sd) were mapped with the BIMA interferometer
\citep{welch96} and the NRAO\footnote{The National Radio Astronomy
  Observatory is a facility of the National Science Foundation,
  operated under cooperative agreement by Associated Universities,
  Inc.} 12m single dish telescope at Kitt Peak.  The SONG sample
galaxies are listed in Table 1.  The details of the survey and
properties of the sample are described in greater detail in
\citet{helfer03}, in particular see \S 2.2).  In some cases the
  distances and position angles used by us are different than those
  noted by \citet{helfer03}; we used the most recent measurements from
  the literature. We note that we checked the results of our study
  with the Helfer et al. distances and position angles and found no
  systematic differences and the results from this study remain
  unchanged.  The selection criteria for the SONG sample were: 
  heliocentric velocity V$_{HEL}$ $ < $ 2000 \kms, declination $\delta
  > $ -20$^{\circ}$, inclination $i < $ 70$^{\circ}$, and apparent
  magnitude B$_T < $ 11.  This lead to a sample that is
  representative of disk galaxies in the local Universe.

The typical data cube has a synthesized beam of
  6$\arcsec$ and a field of view of 3$\arcmin$.  In a 10 \kms\ channel
  the typical noise is $\sim$58 mJy \bm.  The measured fluxes are
  accurate to 15\%; uncertainties are due to the absolute amplitude
  calibration.  The galaxy centers are obtained from the 2MASS Large
  Galaxy Survey \citep{jarrett03}.

With the BIMA SONG data, we measure the molecular gas mass (\mnuc) and
surface density (\snuc) in the central kiloparsec using standard
routines in MIRIAD \citep{sault95}; we discuss why we chose the
central kiloparsec at the end of this section. We use the FCRAO or the
NRAO 12m SONG data for the total molecular gas mass (\mtot) and
surface density (\stot). The mass is calculated using,

\begin{equation}
M_{gas} (M_\sun) = 1.36\ \times \ 1.1 \times 10^{4} \ D_{\rm{Mpc}}^{2}\
S_{CO}\ \rm{(Jy\ km/s) }.
\end{equation}

\noindent The factor of 1.36 is the usual correction for elements
other than hydrogen.  This equation uses a CO-to-H$_{2}$ conversion
factor of 2.8 $\times$ 10$^{20}$ cm$^{-2}$ (K \kms)$^{-1}$
(\citealt{bloeman86}).  The exact value for the conversion factor is
controversial and may vary by factors of 2--4 -- see discussion in
\citet{young91}.  $\Sigma$ is then calculated by dividing this gas
mass by the area.  Twelve of the 44 galaxies have relatively faint
emission in the central kiloparsec; we list a 2$\sigma$ upper limit
for these galaxies.  The results are shown in Tables \ref{derbarprops}
and \ref{derunbarprops} for the barred and unbarred spirals
respectively.  Throughout this paper, we use the RC3 classification
(SB and SAB) for barred spirals.  

As noted earlier, we classify galaxies of type Sa--Sbc as early types,
and galaxies with types Sc--Sd as late types.  Since a significant
number of galaxies in SONG are type Sbc (14 of 44), we discuss briefly
the motivation for the adopted division.  It is based on the
significant differences in the star formation properties between
galaxies of type Sbc and earlier, and those of later types
\citep{ho97b}.  Since star formation depends on gas content, it is
reasonable to use the same boundaries as Ho et al (1997b), otherwise
comparison with previous studies becomes difficult.  Furthermore the
division is also motivated by prior observational and theoretical work
on properties of barred galaxies
\citep{elmelm85,combes93,huang96}. These studies have found a
fundamental difference in the properties of early and late type bars.
Early type bars generally have a ''flat" photometric profile, while
late type bars usually have an ''exponential" profile.  Of the 14 Sbc
galaxies, 11 are barred spirals.  We examined the optical/IR data for
these 11 galaxies, and find that 7 are clearly flat bars (NGC 2903,
NGC 4321, NGC 4303, NGC 3992, NGC 3344, NGC 3953, NGC 4051).  The
optical/IR data for the other four are not good enough to make a
definitive classification, although the CO kinematics in NGC 5005
suggest that it too hosts a strong bar of the type usually found in
early type barred spirals.  For these reasons, we place the Sbc
galaxies in the the early type category.

In column 2 of Tables \ref{derbarprops} and \ref{derunbarprops} we
list the CO flux from the central kiloparsec in the combined (BIMA $+$
NRAO 12m single dish) maps for the 24 galaxies for which we collected
on-the-fly (OTF) data.  In column 3 we list the CO flux from the
BIMA-only data.  In most cases the BIMA-only data recover nearly all the
CO flux.  There are a few exceptions, where the flux in the
combined maps is 40-50\% higher than in the BIMA-only maps:
IC342, NGC 0628, NGC 3521, and NGC 4414.  In these cases, the relative
angular sizes of the beam and galaxy, the inclination, or the CO
distribution likely have contributed to spatial filtering or beam smearing
effects.    

We choose to use the BIMA-only maps to measure \mnuc\ because in most
cases the BIMA-only fluxes are consistent with \mnuc\ from the
combined map (see also Figure 53 in \citealt{helfer03} and discussion
in \citealt{helfer02}).  There is no evidence that barred or unbarred
spirals are preferentially affected by the lack of single-dish data,
and we want to use as homogeneous a dataset as possible.  We did not
correct the BIMA-only data for beam smearing.  Our typical 6$\arcsec$
beam subtends an area of diameter of $<$1 kiloparsec for almost all
galaxies in our sample, as can be seen in Table 2 in \citet{helfer03}
where the area subtended is shown in the plane of the sky.  In the
plane of a galaxy, the beam may subtend a larger region (at most by
cos $i ^{-1}$) along the galaxy minor axis.  For the galaxies in our sample 
this correction is less than a factor of 2 ($i$ $<$ 60$^{\circ}$), and
therefore our beam typically subtends $<$ 1 kpc along the minor
axis of the galaxy plane.

In our analysis we also plot the seven additional galaxies from the
NRO-OVRO survey that are not in common with SONG; these include five
unbarred galaxies, which are rarer in both samples.  We confirmed that
the SONG data are consistent with the NRO-OVRO data by comparing the
measured fluxes for the 13 galaxies in common between the two samples;
after accounting for differences in the CO-to-H$_2$ conversion factor
and adopted distances, the fluxes are consistent within the errors.

We derive \mtot\ from the global CO fluxes obtained by the FCRAO
survey (Table 3 in \citealt{young95}); these fluxes were derived from
fitting an exponential CO profile to several discrete pointings along
the major and minor axis of each galaxy.  SONG, on the other hand,
measured the CO fluxes using OTF maps for 24 galaxies, and discrete
pointings for the remaining galaxies.  The SONG fluxes are consistent
with the FCRAO survey (see Figure 51 in \citealt{helfer03}).  In the
five cases where we did not have FCRAO fluxes, we used the SONG fluxes
for \mtot where available.  These are also listed in Tables
\ref{derbarprops} and \ref{derunbarprops}.

We measured \mnuc\ in the central one kiloparsec because, as discussed
in \S \ref{intro}, most of the evolutionary effects depend on the
accumulation of gas near the nucleus. The choice is also partially
influenced by \citet{sakamoto99a,sakamoto99b} who showed that over
half of the galaxies in their sample have nuclear concentrations that
are well-described by an exponential scale length of less than 500 pc.
Though galaxies in the SONG dataset are less strongly concentrated
(many are not even centrally peaked, see Regan et al. 2001), examining
the inner kiloparsec provides consistency with the previous study. A
different measure of nuclear gas properties might be a dynamical
length scale such as the radius of the largest stable x$_2$ orbit.
However, locating the largest stable x$_2$ orbit requires precise
measurements of the rotation curve and potential \citep{regan03}.
This is difficult near the nuclei of galaxies due to the lack of high
resolution data and the presence of streaming motions induced by the
bar.  Another measure might be to choose a bulge radius as determined
from near-infrared images but deriving a unique bulge-disk
decomposition is difficult.  Yet another measure might be to choose a
distance which is a standard fraction of the galaxy diameter.  This
choice is problematic because there is no evidence that nuclear
properties of galaxies are connected with the disk properties. In
fact, star formation properties of the circumnuclear regions are known
to be distinct from those of the disk \citep{kennicutt98}.  Hence we
choose the central kiloparsec as our fiducial diameter for the nuclear
region.

\section{RESULTS}

\subsection{Molecular Gas Masses in the Central Kiloparsec of Barred and Unbarred Galaxies}\label{mass}

\begin{figure}
\plotone{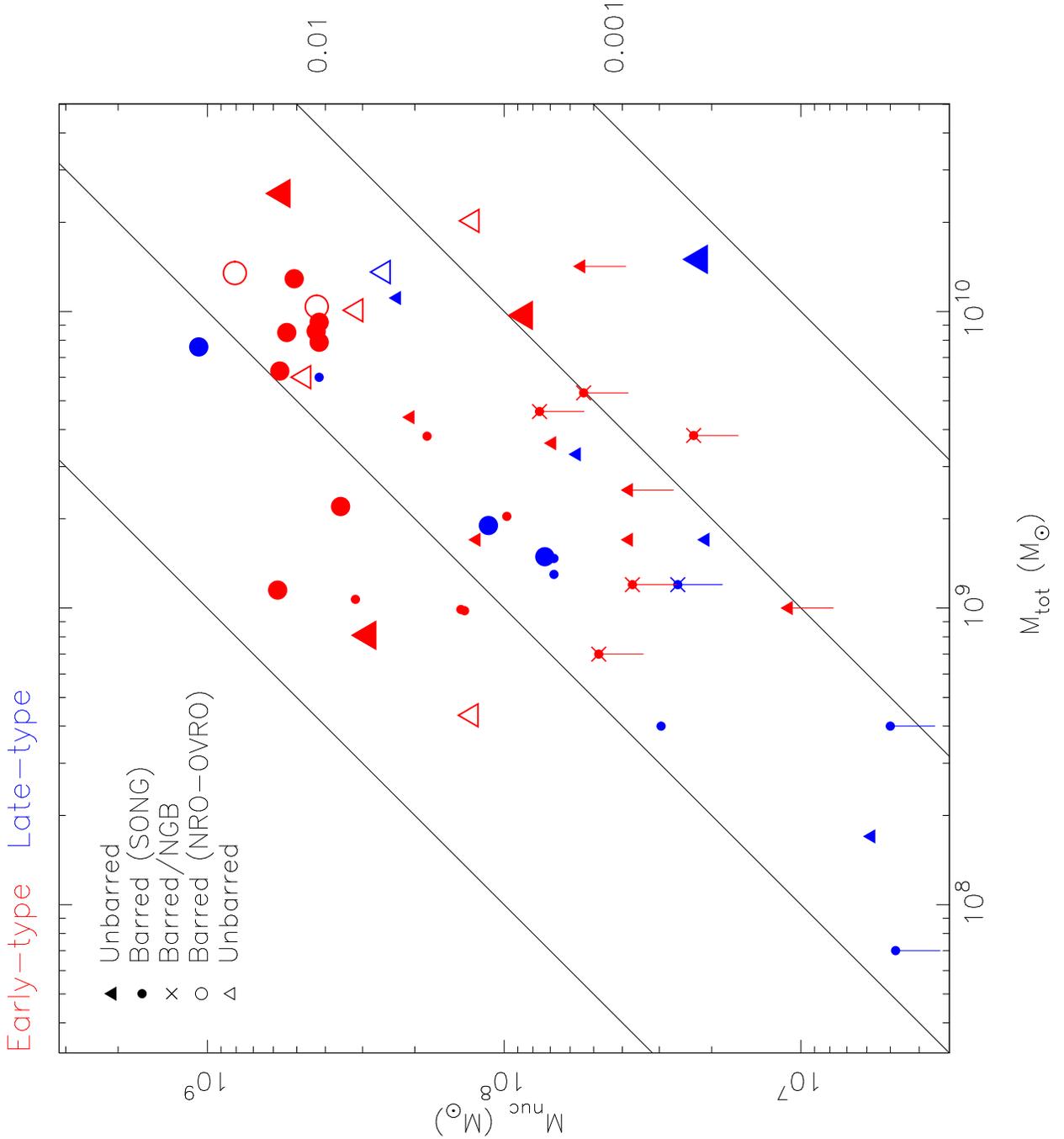}
\caption{Plot of molecular gas mass in solar masses in the central
  kiloparsec (y-axis) and integrated over the entire galactic disk
  (x-axis).  The diagonal lines indicate the molecular gas mass
  fraction in the central kiloparsec.  The barred spirals are
  represented with a circle and the unbarred spirals with a triangle.
  Solid symbols are SONG data whereas open symbols are data from
  galaxies not in common with SONG from \citet{sakamoto99b}. The ``X''
  symbols mark galaxies which have no gas in the bars (NGB).  Galaxies
  with only upper limits are shown with a small vertical segment.
  Larger symbols reflect galaxies which are CO-bright (I$_{CO}>$ 10 K
  \kms\ in an FCRAO pointing). \label{massfig}}
\end{figure}

The mass of the molecular gas in the central kiloparsec (\mnuc) is
plotted against the total molecular gas mass in the galaxy (\mtot) in
Figure \ref{massfig}\footnote{NGC 4699 is not shown in any of the
figures because it only has an upper limit for the single-dish flux.
However it is one of the galaxies with an upper limit for \mnuc\ and
has no gas in its bar.}.  Barred spirals are shown with a circle and
unbarred spirals with a triangle.  SONG galaxies have filled symbols.
The data from \citet{sakamoto99b} for galaxies not in
common with the SONG sample are marked with open symbols.  The
diagonal lines indicate the fraction of the total gas mass in the
central kiloparsec.

Before we discuss the nuclear masses, we note that there is no
obvious difference in the distribution of \mtot\ between barred and
unbarred spirals.  The range for \mtot\ spans $\sim$10$^8$--10$^{10}$
\Msun. For the entire SONG sample, the mean total molecular gas mass,
$<$\mtot$>_{bar}$ = 4$\pm$1 $\times$ 10$^9$\Msun\ for the barred
spirals, and $<$\mtot$>_{unbar}$ = 6$\pm$2 $\times$ 10$^9$\Msun\ for
the unbarred spirals.  Comparing the red symbols (early type) to the
blue symbols (late types) there is also no obvious difference in
\mtot\ values between early and late Hubble type galaxies.  This is
consistent with large single dish surveys of molecular gas in galaxies
(e.g., \citealt{sage93,young95}).

The fraction of the molecular gas that is in the central kiloparsec
varies by almost three orders of magnitude, from 0.001 to nearly 0.6.
Figure 1 shows that galaxies with the highest values of this fraction
and also those with the highest \mnuc\ are mainly early types,
especially bars.  Of the eight galaxies with nuclear gas mass
fractions above 0.1, six are barred, and five of these six are early
Hubble type galaxies.

Another eight barred spirals have only upper limits for their nuclear
molecular gas mass.  Five of these are early type bars; not only is no
gas detected in the central kiloparsec, none is detected within the
entire region interior to the bar.  The existence of early-type barred
spirals without any nuclear gas is consistent with bar-driven gas
transport, as we discuss in the next section. In addition, we note
that these results contrast with the emphasis of previous studies on
large excesses of molecular gas in the central kiloparsec
\citep{sakamoto99b,jogee04}.  For example, \citet{jogee04} conclude
that the central kiloparsec of barred spirals are significantly
different from the outer disks with molecular gas masses in excess of
2$\times$10$^8$ \Msun, up to 3 $\times$10$^9$ \Msun.  However, for the
barred spirals in SONG, we find a range of nuclear masses from
10$^7$--10$^9$ \Msun, with a mean of $\sim$ 2 $\times$ 10$^8$ \Msun.
Indeed, 18 of 29 barred spirals in SONG have \mnuc\ $<$
2$\times$10$^8$ \Msun.  Clearly the high \mnuc\ values emphasized by
previous studies apply only to a subset of barred galaxies.

\subsection{Bar-Driven Gas Transport}\label{sigs}

\begin{figure}
\plotone{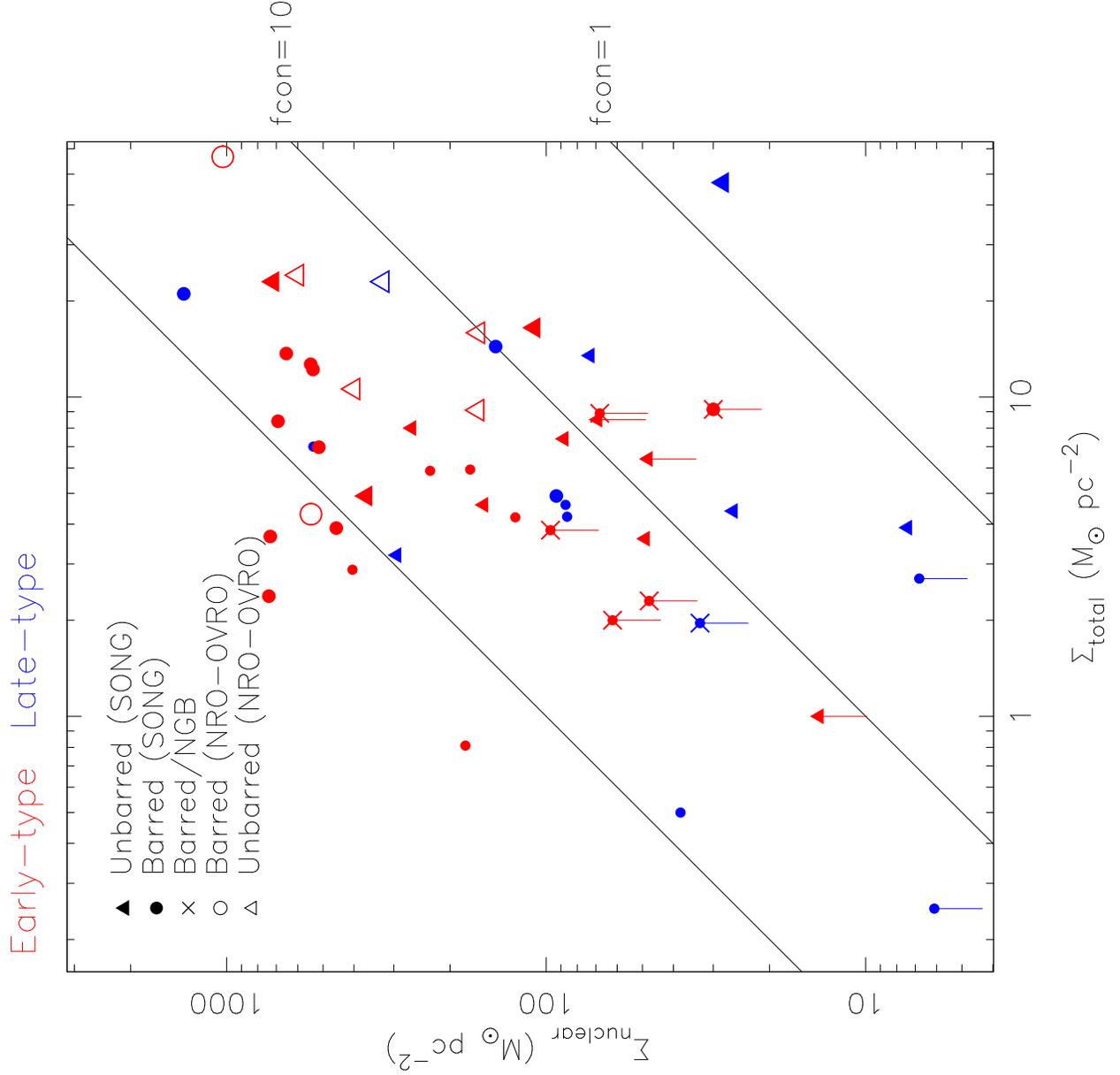}
\caption{Same as Figure \ref{massfig} except the axes now reflect the
  molecular gas surface densities in solar masses per square parsec.  The diagonal lines show the
  concentration index defined as the ratio of the nuclear molecular
  gas surface density to the total molecular gas surface
  density.\label{sigfig}}
\end{figure}

In Figure \ref{sigfig} we plot the molecular gas surface density for
the central kiloparsec (\snuc) versus that averaged over the galaxy
disk inside $D_{25}$ (\stot) for all galaxies shown in Figure
\ref{massfig}.  The diagonal lines now indicate constant values of the
central concentration parameter, \fcon, which we discussed in \S
\ref{intro}.

The figure immediately shows a clear difference in central
concentration (\fcon) between barred and unbarred spirals.  All six
galaxies with \fcon\ $>$ 100 are barred spirals.  At the other end,
for \fcon\ $<$ 10, 7 of the 10 galaxies are unbarred.  The higher
\fcon values in barred spirals thus provide the strongest and most
direct evidence for bar-driven gas transport, consistent with the
results of \citet{sakamoto99b}.  

\subsubsection{Bar Gas Transport in CO-faint Galaxies}

\begin{figure}
\plotone{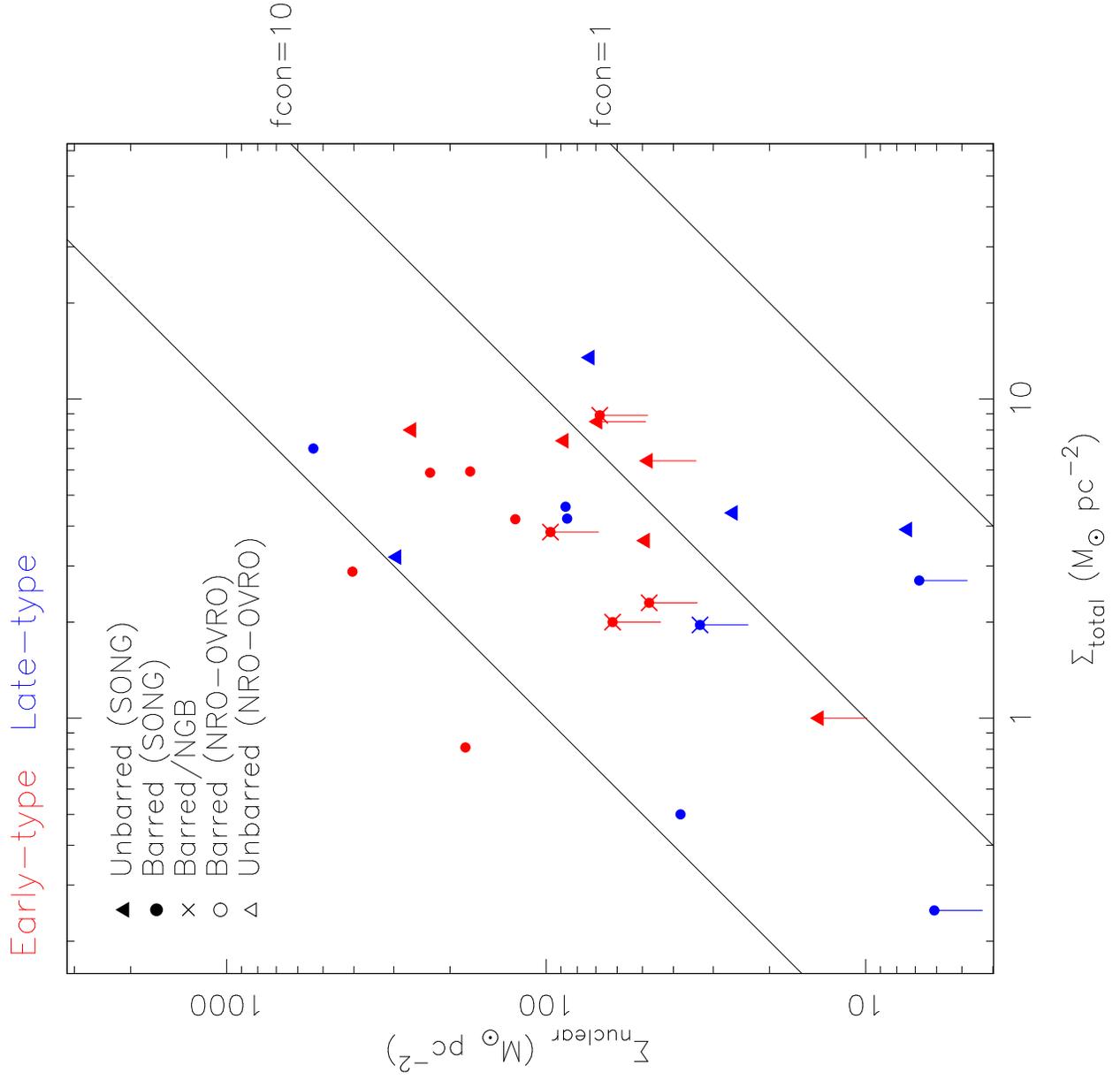}
\caption{Same as Figure \ref{sigfig} except only the CO-faint 
  galaxies are shown. \label{siglowfig}}
\end{figure}

As most of the highest \fcon\ values are in CO-bright galaxies (larger
symbols), we specifically address the question of whether bars also
enhance \fcon\ in CO-faint galaxies.  Figure \ref{siglowfig} shows
only the twenty-six (16 barred and 10 unbarred) SONG galaxies with
FCRAO brightnesses $<$ 10 K \kms.  We see that for \fcon\ $>$ 10, nine
of the 13 galaxies with significant detections in both total and
nuclear surface densities are barred.  Dividing the 26 CO-faint
galaxies into two bins, we see that 11 of the 13 highest
\fcon\ galaxies are barred, compared with only five of the lowest 13
\fcon\ galaxies.  Thus even in the sample of galaxies that are not
CO-bright, we conclude that there is evidence for bar-induced gas
transport.  This evidence is further strengthened upon noting that
most of the low \fcon\ barred galaxies have no gas at all within the bar
co-rotation radius.  The co-rotation radius is believed to be at
1.2$\times$ bar end (e.g., \citealt{a92b}).  Galaxies without gas in
the bar region can be understood as those in which bar driven
accretion has proceeded to the point where all the gas presently
available to the bar has already been driven to the central region and
converted into stars.

Jogee et al (2004b), in a paper that focused on the the role of bars
in starbursts and secular evolution, proposed an evolutionary sequence
for bars and starbursts.  In Type I non-starbursts, gas is still
distributed through the bar region and is only beginning to
accumulate in the central region.  Type II non-starbursts have
significant central concentrations, but the gas has yet to accumulate
to the surface densities required for a starburst.  In starbursts, the 
sufficient gas has accumulated to drive a starburst.

We would therefore add a category, Type III non-starbursts, which are
galaxies in the post-starburst phase.  All the gas within the radius
swept by the bar has already been driven to the nucleus and has been
consumed by a starburst (or sufficient gas has been consumed so that
the current gas surface density is lower than the critical density
needed for a starburst).  It would be interesting to look for evidence
of this in the stellar populations.  Such galaxies will be free of
molecular gas within the bar region until some event enables the
feeding of gas through the barrier that seems to be present near the
bar co-rotation radius.  If this happens, then central gas
accumulations and starbursts would be episodic.

\subsubsection{Bar Gas Transport as a Function of Hubble Type}

\begin{figure}
\plotone{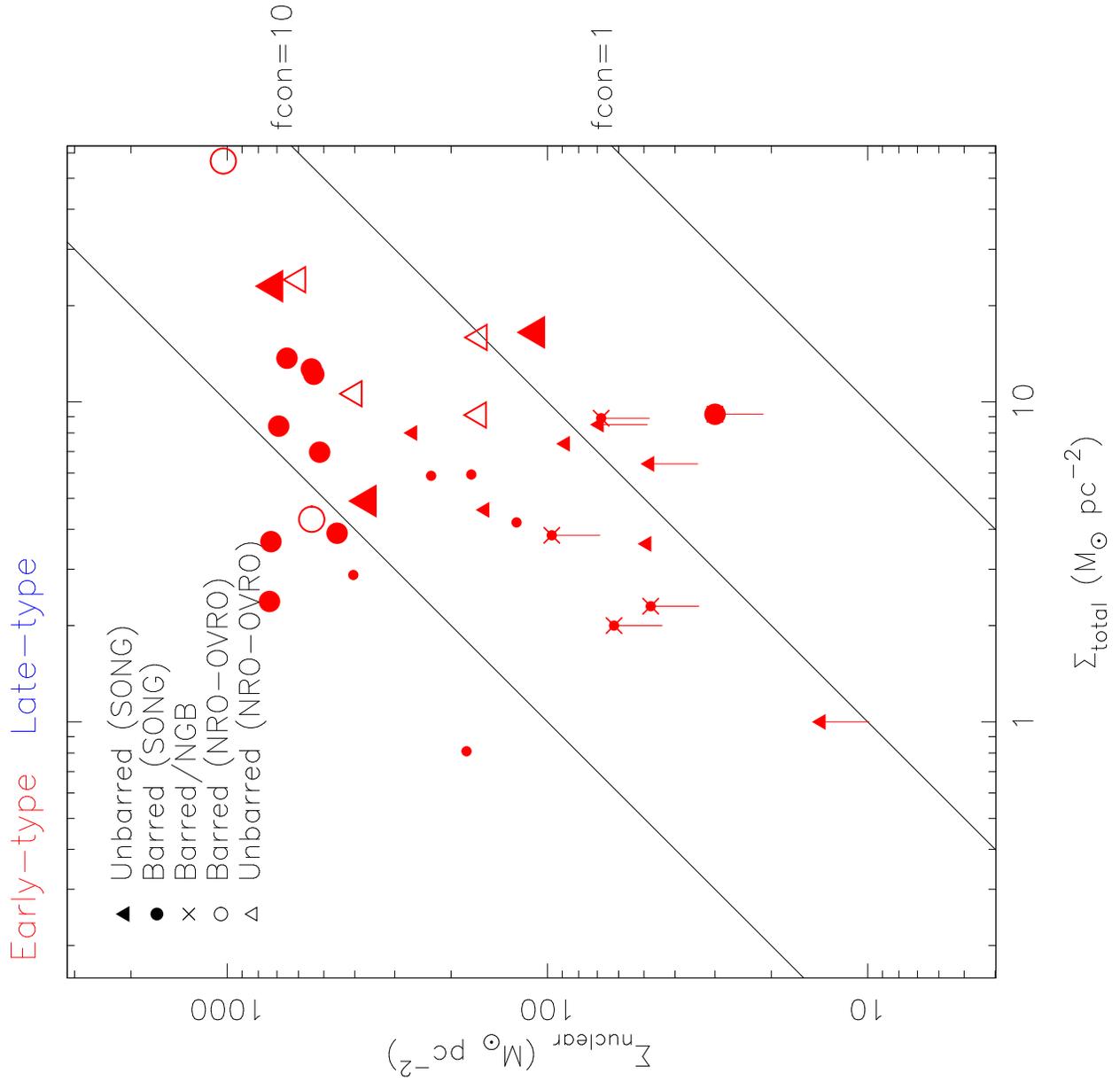}
\caption{Same as Figure \ref{sigfig} except only the early Hubble type (Sa-Sbc)
  galaxies are shown. \label{sigearlyfig}}
\end{figure}

First we consider early Hubble type barred spirals, which we take to
be Hubble types Sbc and earlier.  There are 26 early type galaxies in
SONG, plus six from the Sakamoto sample; these are shown in Figure
\ref{sigearlyfig}.  For galaxies above the median \fcon\ for early
type galaxies, 13 of 16 are barred, compared with only seven of 16
below the median.  The median \fcon\ for the early type bars is 43,
compared with 14 for the unbarred early types.  Clearly, the evidence
for bar driven inflow in early type galaxies is very strong.  

\begin{figure}
\plotone{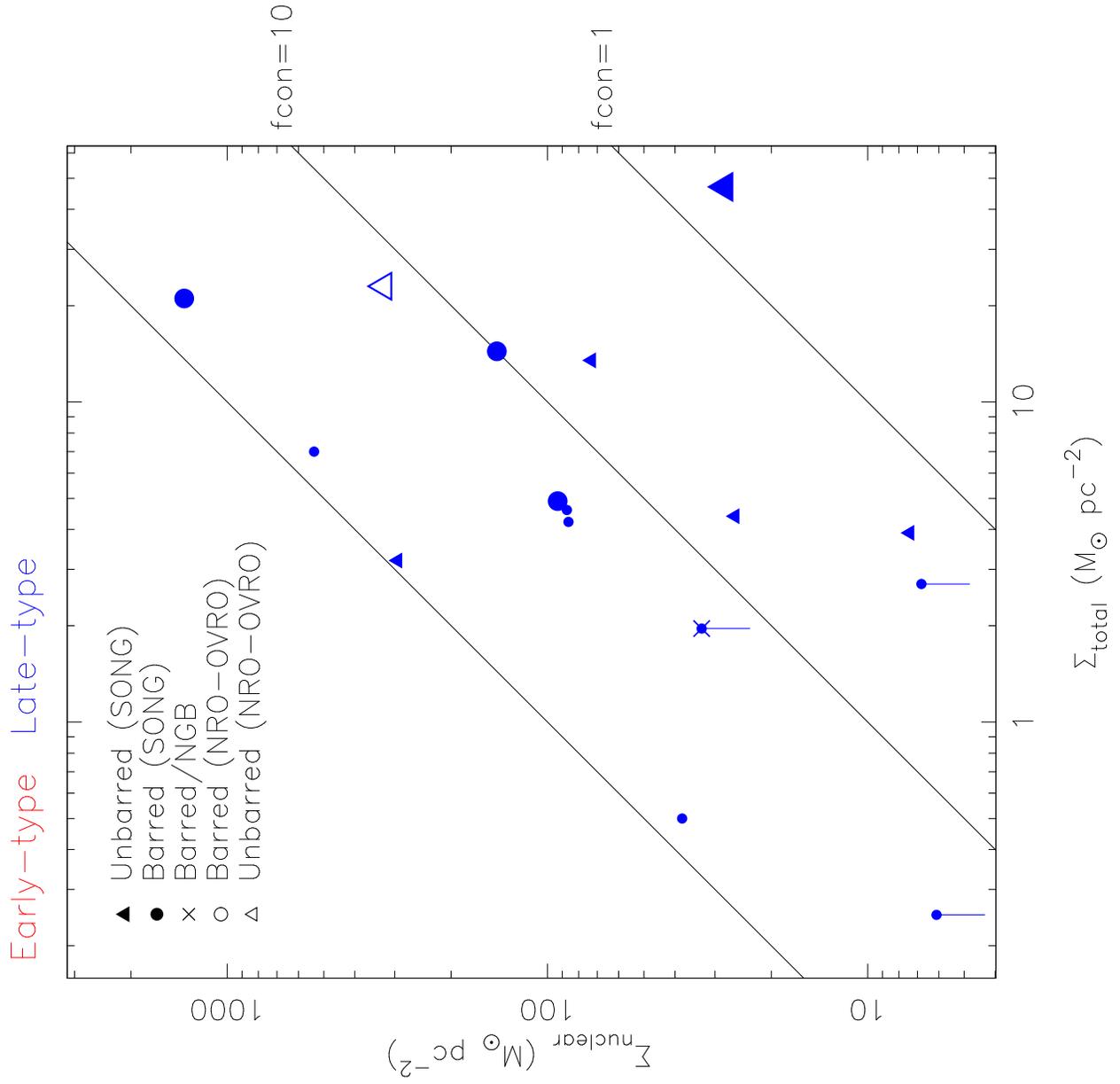}
\caption{Same as Figure \ref{sigfig} except only the late Hubble type (Sc-Sdm)
  galaxies are shown. \label{siglatefig}}
\end{figure}

Next we consider the 16 late type spirals, shown in Figure
\ref{siglatefig}.  For galaxies above the median \fcon, seven of eight
are barred spirals, whereas for galaxies below the median, only three
of eight are barred.  The median \fcon\ for late type bars is 20,
three times higher than that for unbarred late type galaxies.  Thus
there is also good evidence for inflow in late type bars.  We conclude
that bars transport gas inwards regardless of Hubble type.

Although both early type and late type bars have \fcon\ values about
three times higher than their unbarred counterparts, note that the
median \fcon\ for early type barred galaxies is about twice that for
late type bars.  Also, the median \snuc\ (\Msun pc$^{-2}$) for barred
early types is 400, compared to 87 for late type bars.  In part, the
higher \fcon\ and \snuc\ for early types can be understood as a
consequence of the higher critical surface density threshold for star
formation in early types resulting from their steeper rotation
curves. This is probably the reason why even amongst the unbarred
spirals in our sample, the early type galaxies have about a three
times higher median \fcon\ and \snuc\ compared to that for the late
types; however this comparison should be viewed with caution because
of the small number of unbarred galaxies.

Ho et al (1997b) attributed the higher star formation rates in early
type bars to the higher critical density threshold, and predicted the
higher gas surface densities that we indeed find.  However, for early
type bars to {\it maintain} these higher central mass concentrations
requires that the early type bars have significantly higher mass
inflow rates than late type bars.  This is because the gas consumption
rate implied by the star formation rates observed by Ho et al are
higher in early types --- the median values of the star formation rate
are 0.08 compared to 0.007 \Msun yr$^{-1}$ in barred early types to
barred late types, respectively.  That is, not only are the median
surface densities and nuclear masses a factor of four higher in early
type bars, but the gas is consumed at a typical rate more than ten
times higher.  Clearly, the mass inflow rates in early types must be
much higher than for late type bars.

How can we explain this difference?  We believe the reason is related
to the different type of bars that are typically found in early and
late Hubble type galaxies: ``flat'' and ``exponential''
\citep{elmelm85,elmelm96,ohta86,combes93}.  Flat bars, found primarily
in early type galaxies, have a rectangular appearance with flat
shoulders in their photometric profile, are usually associated with
grand-design spiral structure, and lie in the flat part of the rotation
curve.  They are also longer relative to their disks and have higher
amplitudes.  Exponential bars, on the other hand, are located in the
rising part of a rotation curve, and are oval, rather than
rectangular, in appearance. Their photometric profile is exponential,
similar to a galactic disk.  They are shorter relative to their disks
than flat bars, and have a lower amplitude.  Exponential bars are
primarily found in late Hubble type spirals.

Bars in early type galaxies, i.e., flat bars, are predicted to be more
effective at driving gas inwards because longer bars encounter more
gas in the disk, and because the higher non-axisymmetry leads to a
higher inflow rate.  N-body \citep{combes93} and hydrodynamic models
\citep{regan04} predict that weaker bars (fatter bars and/or less
massive bars) drive significantly less mass inward.  The
evidence presented here thus supports the predictions of higher inflow
rates in early type barred spirals.

Another possibility is that the higher CO fluxes in the nuclei of
early type galaxies may reflect a change in the CO emissivity due to
higher pressures and velocity dispersions in the bulge region (see
discussion in \citealt{regan01}).  The increased star formation in the
nuclei of early type spirals may also contribute to an increase in the
pressure.  But if there is indeed less gas at the centers of early
types, then the enhanced star formation implies a higher efficiency
and decreased gas consumption times.  Then the presence of a large
fraction of early type bars with high central concentration of CO
emission is surprising.  A quantitative discussion of this awaits
better extinction-corrected star formation rates.

\subsection{Secular Evolution From Late to Early Type Galaxies?}\label{latetoearly}

Several studies have suggested that galaxies undergo secular evolution
via bar-driven inflows.  One popular evolutionary scenario is the
change of late type galaxies into early types via bar inflows
\citep{friedli95,zhang99}.  We argue that such a transformation is
unlikely.  In order for a late type galaxy to build an early-type
bulge, it would need to add $\sim$10$^{10}$ to $\sim$10$^{11}$
\Msun\ of gas that must then be converted to stars \citep{binney98}.
If we take the best case scenario for such a transformation, i.e. a)
assume that bars are long-lived (10 Gyr), b) transport gas inwards at
sufficiently high rates, c) have an adequate supply of molecular gas,
d) convert enough of the accumulated gas to stars, and e) scatter the
stars in the vertical dimension to build a bulge, then a late type
galaxy can build an early type bulge.  Some of these conditions are
difficult, if not impossible to reconcile with observational data.

While the high bar fraction at z $>$ 0.7 \citep{sheth03} suggests that
bars are likely to be long-lived, it is unlikely that the mass inflow
rates or the gas reservoir in late Hubble type galaxies are sufficient
to build an early type bulge.  Based on the differences in
\fcon\ values, we noted that the mass inflow rate in late Hubble type
galaxies is lower than in early Hubble types.  The inflow rate in the
NRO-OVRO sample was estimated by \citet{sakamoto99b} to be $\sim$0.1-1
\Msun\ \yr.  In later type galaxies the rate would be even lower.  We
note that at the median nuclear star formation rate for late type bars
measured by Ho et al. (1997b) data (0.007 \Msun yr$^{-1}$) it would
take $10^{12}$ yr to build even a $10^{10}$\Msun\ bulge.  At the upper
end of gas consumption rates for late type bars (1 \Msun yr$^{-1}$),
it would be possible to build a small bulge in $10^{10}$ yr; however,
the data demonstrate that this is certainly not typical.  

Moreover, late type bars are small and do not have access to a vast
molecular gas reservoir.  In SONG, the average length of a late type
bars is 3 kpc (the average length of bars in the entire sample is 5
kpc, \citealt{sheth01}).  To measure the available gas reservoir, let
us assume this length and the average \stot\ for SONG (8
\Msun\ \ppcsq). Then the available gas reservoir for a typical late
type spiral is only 5$\times$ 10$^7$ \Msun.  This value may be higher
because the average \stot\ is a measurement over the entire disk.  In
the bar region, the available gas reservoir may be higher.  But even
if we assume a $\Sigma$ that is ten times higher, i.e., 80 \Msun
\ppcsq\, the gas reservoir would still be too small.  It is possible
that accretion from mergers or gas transport from outer parts of the
galactic disk may increase this number but it is at least two orders
of magnitude lower than that required for creating an early-type bulge
in a late type galaxy.  {\it Thus we believe that it is unlikely that a
late type galaxy evolves into an early types, consistent with
\citet{kormendy04}}.

This, however, does not imply that there is no secular evolution in
barred spiral galaxies.  To the contrary, the evidence presented here
for bar-driven gas inflows, combined with the enhanced star formation
activity in barred spirals, indicates that the central regions of
barred galaxies undergo significant changes.  They are probably the
most important changes in the evolution of disks since z$\sim$1 when
the merging activity began to decline
\citep{baugh96,ferguson00}. Another possible secular evolutionary
sequence could be the dissolution of bars by an increasing central
concentration of mass.  Models have debated the exact impact of
increasing central concentration (e.g.,
\citealt{friedli93,norman96,bournaud02,athanassoula03,shen04}).
Tentative evidence for such dissolution was presented in \citet{das03}
who showed a correlation between the bar ellipticity and the central
mass concentration.  

The data presented here lend credence to the formation of
pseudo-bulges in barred spirals; pseudo-bulges and secular evolution
via gas inflows has recently been reviewed by \citet{kormendy04}.  In
this context, we particularly point out the significant population of
early-type barred spirals (6 of 19) that have no molecular gas
detected in nuclear region, and very little detected in the region
within the bar co-rotation radius.  This suggests that the gas inside
the bar co-rotation radius has already been driven into the nuclear
region, where it has been consumed by star formation.  Amongst the
late type spirals, there are three barred galaxies with upper limits
to the molecular gas in the central kiloparsec.  One of them has no
gas within the bar co-rotation radius.  As already discussed, the
process of gas accretion is slower in late types.  The star formation
rates in late types are also lower because late type spirals have
lower \scrit, the critical gas surface densities for star formation
(e.g., \citealt{toomre64,kennicutt89}).  So it is not unusual to find
fewer examples of late type barred spirals with no gas in the center
and bar regions.  The secular evolutionary sequence of gas inflow and
subsequent star formation simply occurs at a slower pace in these
galaxies; over time, even the late type spirals will build
pseudo-bulges.  Signatures of post-starburst populations in barred
spirals with low \mnuc\ and little gas in the bar region would provide
further confirmation of this evolutionary sequence.

Another way to interpret the galaxies without gas in the center is
that a third of the barred galaxies, whether they are early-type or
late-type, are in a quiescent state, i.e. without measurable gas in
the center or the bar region.  This indicates that if the fueling via
bars is periodic, then bars are ``actively'' fueling the center
two-thirds of the time.  When better measurements of gas inflow are
obtained, this duty cycle will need to be taken into account to
measure the impact of the gas inflow over the lifetime of the bar and
galaxy.

\section{Conclusions}\label{conclusions}

With a larger and more representative sample of nearby galaxies from
SONG, we compared the central concentration of molecular gas in both
CO-bright and CO-faint spiral galaxies spanning a range of Hubble types.  In
all cases we found clear evidence of more centrally concentrated
molecular gas distributions in barred spirals.  This is the strongest
and most conclusive evidence for the bar-driven transport of molecular
gas to the central kiloparsec of galaxies.

We find that late Hubble types barred spirals are less centrally
concentrated than early Hubble types.  This, coupled with the enhanced
star formation activity observed in early type bars, indicates higher
mass accretion rates in early Hubble type spirals.  The differences
are probably related to the longer and stronger ``flat'' bars that are
found preferentially in early Hubble type galaxies.  
The observation of a significant subset of early type barred spirals
with little or no gas within the bar region is consistent with higher
accretion rates; these galaxies have the expected conditions of a
post-starburst galaxy, where the large stockpiles of gas driven inward
by the bar have already been converted into stars.

Contrary to previous suggestions, the evidence indicates that it is
highly unlikely for a late type galaxy to evolve into an early type
via bar-induced gas inflow.  Nonetheless, secular evolutionary
processes are undoubtedly present, and pseudo-bulges are inevitable
because of the bar-induced gas inflow.  Evidence for pseudo-bulges is
likely to be clearest in early type galaxies because of their
higher inflow rates and higher star formation activity.

\acknowledgments

The authors are indebted to the referee for the detailed and thorough
examination of this manuscript and the many helpful comments and
suggestions that s/he provided.  This work would not have been
possible without the rest of the SONG team members (T. Helfer,
T. Wong, L. Blitz and D. Bock) and the dedicated observatory staff at
Hat Creek and at the Laboratory for Millimeter-wave Astronomy at the
University of Maryland.  We thank T. Beasley, D.M. Elmegreen,
B.G. Elmegreen, A. Harris, J. Kenney, J. Knapen, N. Scoville,
E. Schinnerer, I. Shlosman, L.E. Strubbe, and T. Treu for invaluable
and insightful discussions.

\clearpage

\clearpage

\clearpage

\clearpage

\clearpage
%

\begin{landscape}
\begin{deluxetable}{rrrrrrrrll}
\tabletypesize{\scriptsize}
\tablecaption{BIMA SONG Sample\label{sampletab}}
\tablehead{\colhead{Galaxy}& \colhead{$\alpha$(J2000)r\tablenotemark{a}} &\colhead{$\delta$(J2000)r\tablenotemark{a}}& \colhead{D\tablenotemark{c}$_{25}$}&\colhead{PA\tablenotemark{d}}&\colhead{$i$\tablenotemark{e}}&\colhead{V\tablenotemark{f}$_{Hel}$}&\colhead{D\tablenotemark{g}}&\colhead{Type} &\colhead{OTF\tablenotemark{i}} \\
\colhead{} & \colhead{(h m s)} & \colhead{($^{\circ}$ $\arcmin$ $\arcsec$)} &  \colhead{($\arcmin$)} & \colhead{($\deg$)} & \colhead{($\deg$)} &\colhead{(\kms)} &\colhead{(Mpc)} & \colhead{(RC3)} &\colhead{Data}}
\startdata
& & & & Barred & & & & & \\
\tableline
NGC 0925  & 02 27 16.9 & +33 34 44.0 & 10.5 & 102 & 55.8 & 553 &  9.3$^k$  &  SAB(s)d  &    ... \\
IC 342$^j$   & 03 46 48.5 & +68 05 46.0 & 21.4 &  37$^{zz}$  &  31.0$^{zz}$  & 34 &  2.1l  &  SAB(rs)cd  &    Y \\
NGC 2403  & 07 36 51.3 & +65 36 09.2 & 21.9 & 127 & 55.8 & 131 &  2.9$^m$  &  SAB(s)cd  &    ... \\
NGC 2903$^j$  & 07 36 51.3 & +65 36 09.2 & 12.6 & 17 & 61.4 & 556 &  7.3$^n$  &  SAB(rs)bc  &    Y \\
NGC 3184  & 10 18 16.9 & +41 25 27.8 & 7.4 & 135 & 21.1 & 592 &  8.7$^o$  &  SAB(rs)cd  &    ... \\
NGC 3344  & 10 43 31.1 & +24 55 20.0 & 7.1 & 24 & 25.5 & 586 &  12.5$^p$  &  (R)SAB(r)bc  &    ... \\
NGC 3351  & 10 43 57.7 & +11 42 13.0 & 7.4 & 13 & 47.8 & 778 &  10.1$^q$  &  SB(r)b  &    Y \\
NGC 3368$^j$  & 10 46 45.7 & +11 49 11.8 & 7.6 & 5 & 46.2 & 897 &  11.2$^r$  &  SAB(rs)ab   &  ... \\ 
NGC 3521  & 11 05 48.5 & -00 02 09.2 & 11 & 163 & 62.1 & 805 &  7.2$^s$  &  SAB(rs)bc  &   Y \\ 
NGC 3627  & 11 20 15.0 & +12 59 28.6 & 9.1 & 173 & 62.8 & 727 &  11.1$^t$  &  SAB(s)b  &  Y \\
NGC 3726  & 11 33 21.1 & +47 01 44.7 & 6.2 & 10 & 46.2 & 849 &  11.7$^u$  &  SAB(r)c  &    ... \\
NGC 3953  & 11 53 49.0 & +52 19 35.6 & 6.9 & 13 & 59.9 & 1054 &  14.1$^v$  &  SB(r)bc  &   ... \\
NGC 3992  & 11 57 35.9 & +53 22 28.3 & 7.6 & 68 & 51.9 & 1048 &  14.2$^w$  &  SB(rs)bc  &  ... \\
NGC 4051  & 12 03 09.6 & +44 31 52.5 & 5.3 & 135 & 42.2 & 725 &  9.4$^x$  &  SAB(rs)bc  &   ... \\
NGC 4258  & 12 18 57.6 & +47 18 13.4 & 18.6 & 150 & 67.1 & 448 &  7.3$^y$  &  SAB(s)bc  &  Y \\
NGC 4303  & 12 21 54.9 & +04 28 24.9 & 6.5 & 7 & 27 & 1566 &  15.2$^z$  &  SAB(rs)bc  &    Y \\
NGC 4321$^j$  & 12 22 54.8 & +15 49 20.6 & 7.4 & 30 & 31.7 & 1571 &  16.1$^{aa}$  &  SAB(s)bc  & Y \\ 
NGC 4490  & 12 30 36.3 & +41 38 37.1 & 6.3 & 125 & 60.7 & 565 &  7.7$^{ab}$  &  SB(s)d-pec  &    ... \\ 
NGC 4535  & 12 34 20.3 & +08 11 51.9 & 7.1 & 0 & 44.9 & 1961 &  16.0$^{ac}$  &  SAB(s)c  &    ... \\
NGC 4548  & 12 35 26.4 & +14 29 46.8 & 5.4 & 150 & 37.4 & 486 &  15.9$^{ad}$  &  SB(rs)b  &  ... \\
NGC 4559  & 12 35 57.6 & +27 57 35.1 & 10.7 & 150 & 66 & 815 &  9.7$^{ae}$  &  SAB(rs)cd  &    ... \\
NGC 4569$^j$  & 12 36 49.8 & +13 09 46.3 & 9.6 & 23 & 62.8 & -235 &  16.8$^{af}$  &  SAB(rs)ab  &  Y \\
NGC 4579  & 12 37 43.5 & +11 49 05.1 & 5.9 & 95 & 37.4 & 1519 &  16.8$^{ag}$  &  SAB(rs)b  &   ... \\ 
NGC 4699  & 12 49 02.18 & -08 39 51.4 & 3.8 & 45 & 46.2 & 1427 &  19.0$^{ah}$  &  SAB(rs)b  &  ... \\
NGC 4725  & 12 50 26.6 & +25 30 02.7 & 10.7 & 35 & 44.9 & 1206 &  12.6$^{ai}$  &  SAB(r)ab-pec  &  ... \\ 
NGC 5005$^j$  & 12 50 26.6 & +25 30 02.7 & 5.8 & 65 & 61.4 & 946 &  21.3$^{aj}$  &  SAB(rs)bc  &  Y \\
NGC 5248$^j$  & 13 37 32.0 & +08 53 06.2 & 6.2 & 110 & 43.6 & 1153 &  22.7$^{ak}$  &  SAB(rs)bc  &    Y \\
NGC 5457  & 14 03 12.5 & +54 20 55.5 & 28.8 & 35 & 21.1 & 241 &  6.5$^{al}$  &  SAB(rs)cd  &   Y \\
NGC 6946$^j$  & 20 34 52.3 & +60 09 13.2 & 11.5 & 35 & 31.7 & 48 &  6.4$^{am}$  &  SAB(rs)cd  &    Y \\
\tableline
& & & & Unbarred & & & & & \\
\tableline
NGC 0628  & 01 36 41.7 & +15 47 00.5 & 10.5 & 25 & 24.2 & 657 &  7.3$^{an}$  &  SA(s)c  &   Y \\
NGC 1068  & 02 42 40.7 & -00 00 47.8 & 7.1 & 70 & 31.7 & 1136 &  18.0$^{ao}$  &  (R)SA(rs)b   &  Y \\ 
NGC 2841  & 09 22 02.6 & +50 58 35.3 & 8.1 & 147 & 64.1 & 638 &  9.5$^{ap}$  &  SA(r)b  &  ... \\
NGC 2976  & 09 47 15.4 & +67 54 59.0 & 5.9 & 143 & 62.8 & 3 &  4.3$^{aq}$  &  SAc-pec  &    ... \\
NGC 3031  & 09 55 33.1 & +69 03 54.9 & 26.9 & 157 & 58.4 & -34 &  3.3$^{ar}$  &  SA(s)ab  &  ... \\ 
NGC 3938  & 11 52 49.4 & +44 07 14.6 & 5.4 & 20 & 14 & 809 &  11.3$^{as}$  &  SA(s)c  &   Y \\
NGC 4414$^j$  & 12 26 27.08 & +31 13 24.7 & 3.6 & 155 & 55.8 & 716 &  19.1$^{at}$  &  SA(rs)c?  &  Y \\ 
NGC 4450  & 12 28 29.6 & +17 05 05.8 & 5.3 & 175 & 42.2 & 1954 &  16.0$^{au}$  &  SA(s)ab  &  ... \\
NGC 4736$^j$  & 12 50 53.1 & +41 07 12.5 & 11.2 & 105 & 35.6 & 308 &  6.6$^{av}$  &  (R)SA(r)ab  &  Y \\
NGC 4826$^j$  & 12 56 43.6 & +21 40 57.6 & 10 & 115 & 57.5 & 408 &  5.0$^{aw}$  &  (R)SA(rs)ab  &  Y \\
NGC 5033  & 13 13 27.5 & +36 35 37.1 & 10.7 & 170 & 62.1 & 875 &  21.3$^{ax}$  &  SA(s)c  &  Y \\
NGC 5055$^j$  & 13 15 49.3 & +42 01 45.4 & 12.6 & 105 & 54.9 & 504 &  7.2$^{ay}$  &  SA(rs)bc &  Y \\
NGC 5194$^j$  & 13 29 52.6 & +47 11 42.9 & 11.2 & 163 & 51.9 & 463 &  8.4$^{az}$  &  SA(s)bc-pec &  Y \\ 
NGC 5247  & 13 38 03.0 & -17 53 02.5 & 5.6 & 20 & 29.4 & 1357 &  15.2$^{bb}$  &  SA(s)bc  &  Y \\
NGC 7331  & 22 37 04.0 & +34 24 57.3 & 10.5 & 171 & 69.2 & 821 &  15.1$^{bc}$  &  SA(s)b  &  Y \\

\enddata
\tablenotetext{a}{2MASS Galaxy Centers from the LGA (Jarett et al. 2003)}
\tablenotetext{c}{Optical diameter (RC3)}
\tablenotetext{d}{Position angle (RC3)}
\tablenotetext{e}{Inclination (RC3)}
\tablenotetext{f}{Heliocentric velocity (NED)}

\tablenotetext{g}{Adopted distance, k:\citet{sohn98};
l:\citet{karachentsev93}; m:\citet{metcalfe91}; n:\citet{planesas97};
o:\citet{pompei97}; p:\citet{corbelli81}; q:\citet{graham97};
r:\citet{tanvir99}; s:\citet{thornley96}; t:\citet{saha99};
u:\citet{vanderkruit71}; v:RC3, using H$_o$=75; w:\citet{gottesman82};
x:\citet{liszt95}; y:\citet{herrnstein}; z:\citet{molla99};
aa:\citet{ferrarese96}; ab:\citet{tully88}; ac:\citet{macri99};
ad:\citet{graham99}; ae:\citet{vogler96}; af:\citet{barth98};
ag:\citet{ho99a}; ah:\citet{bower93}; ai:\citet{gibson99};
aj:\citet{barth98}; ak:\citet{tully88}; al:\citet{stetson98};
am:\citet{sharina97}; an:\citet{sharina96}; ao:lit. values vary from
14-22 Mpc, assumed 18 Mpc as mid-point value; ap:\citet{block99};
aq:\citet{karachentsev91}; ar:\citet{paturel98}; as:\citet{jiminez99};
at:\citet{turner98}; au:\citet{tully88}; av:\citet{mulder93};
aw:\citet{rubin94}; ax:\citet{ho99b}; ay:\citet{tully88};
az:\citet{feldmeir97}; bb:\citet{grosbol98}; bc:\citet{hughes98}}

\tablenotetext{i}{Indicates whether On-The-Fly data was collected as part of SONG.}
\tablenotetext{j}{Galaxies in common between SONG and NRO-OVRO sample \citep{sakamoto99b}}
\tablenotetext{zz}{From kinematic fitting, \citet{cros00}}

\end{deluxetable}
\end{landscape}


\clearpage


%

\begin{landscape}
\begin{deluxetable}{lrrrrrrrrr}
\tabletypesize{\scriptsize}
\tablecaption{Derived Properties for Barred SONG galaxies\label{derbarprops}}
\tablehead{\colhead{Galaxy}& \colhead{S$^{CO}_{nuc}$\tablenotemark{a}} &\colhead{S$^{CO}_{nuc}$\tablenotemark{a}} & \colhead{Rms} & \colhead{M$_{nuc}$\tablenotemark{b}} &\colhead{$\Sigma_{nuc}$\tablenotemark{c}} & \colhead{S$^{CO}_{tot}$\tablenotemark{d}}& \colhead{M$_{tot}$\tablenotemark{e}} &\colhead{$\Sigma_{tot}$\tablenotemark{f}} &\colhead{$\Sigma_{nuc}/\Sigma_{tot}$\tablenotemark{g}} \\
\omit & \colhead{COMB} &\colhead{BIMA} &\colhead{in BIMA} & \omit & \omit & \omit & \omit & \omit & \omit \\
\omit & \colhead{(Jy \kms)} &\colhead{(Jy \kms)} &\colhead{(Jy \kms)} & \colhead{(10$^7$\Msun)} & \colhead{(\Msun\ pc$^{-2}$)} &\colhead{(Jy \kms)} &\colhead{(10$^8$\Msun)} & \colhead{(\Msun\ pc$^{-2}$)} &}
\startdata
& & & & Early Types & & & & & \\
\tableline
NGC 2903 &  462 &  447  &  13  &  35.6  &  454  &  2740  &  22.0 &  3.89  &  117 \\
NGC 3344$^u$ &  ... &  2  &  8  &  3.70$^u$  &  47.6$^u$  &  520  &  12.0 &  2.30  &  20.7$^u$ \\
NGC 3351 &  207 &  208  &  9  &  31.7  &  404  &  700  &  10.7 &  2.88  &  140 \\
NGC 3368 &  ... &  309  &  11  &  58.0  &  738  &  610  &  11.5 &  2.38  &  310 \\
NGC 3521$^{u}$ &  53 &  4  &  15  &  2.30$^{u}$  &  30.0$^{u}$  &  4920  &  38.2 &  9.15  &  3.28$^{u}$ \\
NGC 3627 &  244 &  233  &  7  &  43.0  &  546  &  4660  &  85.8 &  12.7  &  43.1 \\
NGC 3953$^u$ &  ... &  7  &  9  &  5.40$^u$  &  68.0$^u$  &  1790  &  53.2 &  8.90  &  8.04$^u$ \\
NGC 3992$^u$ &  ... &  6  &  8  &  4.80$^u$  &  62.0$^u$  &  ...  &  7.00$^s$ &  2.00$^s$  &  31.0$^s$ \\
NGC 4051 &  ... &  103  &  3  &  13.6  &  173  &  740  &  9.80 &  5.93  &  29.2 \\
NGC 4258 &  180 &  176  &  13  &  14.0  &  179  &  1240  &  9.90 &  0.81  &  221 \\
NGC 4303 &  135 &  122  &  4  &  42.0  &  537  &  2280  &  78.8 &  12.2  &  44.0 \\
NGC 4321 &  140 &  132  &  5  &  51.0  &  651  &  3340  &  129 &  13.7  &  47.5 \\
NGC 4548 &  ... &  26  &  5  &  9.80  &  125  &  540  &  20.4 &  4.20  &  29.8 \\
NGC 4569 &  141 &  136  &  5  &  57.0 &  731  &  1500  &  63.0 &  3.66  &  200 \\
NGC 4579 &  ... &  43  &  8  &  18.2 &  231  &  910  &  38.0 &  5.88  &  39.3 \\
NGC 4699$^u$ &  ... &  3  &  6  &  6.50$^u$  &  83.0$^u$  &  ...  &  ... &  ...  &  ... \\
NGC 4725$^u$ &  ... &  15  &  16  &  7.60$^u$  &  97.0$^u$ &  1950  &  46.0 &  3.83  &  25.3$^u$ \\
NGC 5005 &  ... &  80  &  2  &  54.0  &  691  &  1260  &  85.0 &  8.40  &  82.3 \\
NGC 5248 &  57 &  54  &  2  &  42.0  &  515  &  1190  &  92.0 &  6.97  &  76.0 \\
\tableline
& & & & Late Types & & & & & \\
\tableline
IC 342 &  2379 &  1713  &  75 &  11.3  &  144  &  29220  &  19.0 &  14.4  &  10.0 \\
NGC 0925$^u$ &  ... &  6  &  10  &  2.60$^u$  &  33.0$^u$  &  960  &  12.0 &  1.96  &  16.8$^u$ \\
NGC 2403$^u$ &  ... &  15  &  19  &  0.50$^u$  &  6.10$^u$  &  540  &  0.70 &  0.25  &  24.4$^u$ \\
NGC 3184 &  ... &  60  &  7  &  6.80  &  87.0  &  1120  &  13.0 &  4.60  &  18.9 \\
NGC 3726 &  ... &  33  &  3  &  6.80  &  86.0  &  720  &  14.7 &  4.22  &  20.4 \\
NGC 4490$^u$ & ... &  0  &  3  &  0.50$^u$  &  6.80$^u$  &  480  &  4.00 &  2.70 &  2.52$^u$ \\
NGC 4535 &  ... &  110  &  5  &  42.0  &  536  &  1570  &  60.0 &  7.00  &  76.6 \\
NGC 4559 &  ... &  21  &  9  &  2.96  &  38.0 &  ...  &  4.00$^s$ &  0.50$^s$  & 76.0$^s$ \\
NGC 5457 &  123 &  116  &  7  &  7.30  &  93.0  &  ...  &  14.9$^s$ &  4.90$^s$  &  19.0 \\
NGC 6946 &  1691 &  1747  &  15  &  107  &  1363  &  12370  &  76.0 &  21.1  &  64.7 \\ 
\enddata
\tablenotetext{a}{CO flux in central kiloparsec measured from BIMA SONG}
\tablenotetext{b}{Total gas mass in the central kiloparsec calculated from the SONG flux. All masses are 
calculated using M = 1.36 M(H$_2$), M(H$_2$)=1.1$\times$10$^4$ D$^2$
S$_{CO}$ where D is in Mpc and S$_{CO}$ is in Jy \kms.  We use a
CO-to-H$_2$ coversion factor of 2.8$\times$10$^{20}$ cm$^{-2}$ (K
\kms))$^{-1}$, 7\% smaller than \citet{sakamoto99b}.}
\tablenotetext{c}{Nuclear gas surface density derived by dividing the mass by $\pi$ (500 pc)$^2$. }
\tablenotetext{d}{Total CO flux from galaxy (disk+nucleus) from the FCRAO survey \citep{young95}}
\tablenotetext{e}{Total gas mass in the galaxy (disk + nucleus)
calculated from the FCRAO flux}
\tablenotetext{f}{Total gas surface density in the galaxy (disk +
nucleus) calculated from the FCRAO flux}
\tablenotetext{g}{Ratio of the nuclear gas surface density to the disk
gas surface density, defined as f$_{con}$ in \S \ref{intro}}
\tablenotetext{s}{Using galaxy-averaged flux from SONG data instead of the FCRAO survey.}
\tablenotetext{u}{2$\sigma$ upper limit}
\end{deluxetable}
\end{landscape}


\clearpage
\begin{landscape}
\begin{deluxetable}{lrrrrrrrrr}
\tabletypesize{\scriptsize}
\tablecaption{Derived Properties for Unbarred SONG galaxies\label{derunbarprops}}
\tablehead{\colhead{Galaxy}& \colhead{S$^{CO}_{nuc}$\tablenotemark{a}} &\colhead{S$^{CO}_{nuc}$\tablenotemark{a}} & \colhead{Rms} & \colhead{M$_{nuc}$\tablenotemark{b}} &\colhead{$\Sigma_{nuc}$\tablenotemark{c}} & \colhead{S$^{CO}_{tot}$\tablenotemark{d}}& \colhead{M$_{tot}$\tablenotemark{e}} &\colhead{$\Sigma_{tot}$\tablenotemark{f}} &\colhead{$\Sigma_{nuc}/\Sigma_{tot}$\tablenotemark{g}} \\
\omit & \colhead{COMB} &\colhead{BIMA} &\colhead{in BIMA} & \omit & \omit & \omit & \omit & \omit & \omit \\
\omit & \omit &\colhead{(Jy \kms)} & \omit & \colhead{(10$^7$\Msun)} & \colhead{(\Msun\ pc$^{-2}$)} &\colhead{(Jy \kms)} &\colhead{(10$^8$\Msun)} & \colhead{(\Msun\ pc$^{-2}$)} &}

\startdata
& & & & Early Types & & & & & \\
\tableline
NGC 1068 & 120 & 116 & 9 & 56.0 & 716 & 5160 & 250 & 23.0 & 31.1 \\
NGC 2841$^u$ & ... & -11 & 14 & 3.80$^u$ & 48.0$^u$ & 1870 & 25.0 & 6.40 & 7.50$^u$ \\
NGC 3031 & ... & 37 & 34 & 1.10$^u$ & 14.0$^u$ & ... & 10.0$^s$ & 1.00$^s$ & 14.0$^s$ \\
NGC 4450 & ... & 12 & 5 & 3.80 & 49.0 & 450 & 17.0 & 3.60 & 13.6 \\
NGC 4736 & 239 & 190 & 15 & 12.4 & 157 & 2560 & 17.0 & 4.60 & 34.1 \\
NGC 4826 & 760 & 769 & 24 & 28.8 & 366 & 2170 & 8.10 & 4.90 & 74.7 \\
NGC 5055 & 302 & 267 & 13 & 20.7 & 264 & 5670 & 44.0 & 8.00 & 33.0 \\
NGC 5194 & 103 & 81 & 11 & 8.55 & 109 & 9210 & 97.0 & 16.5 & 6.61 \\
NGC 5247 & 29 & 20 & 3 & 6.90 & 88.0 & 1030 & 36.0 & 7.40 & 11.9 \\
NGC 7331$^u$ & 4 & -3 & 8 & 5.50$^u$ & 69.0$^u$ & 4160 & 142 & 8.50 & 8.12$^u$ \\
\tableline
& & & & Late Types & & & & & \\
\tableline
NGC 0628 & 41 & 26 & 8 & 2.10 & 26.0 & 2160 & 17.0 & 4.40 & 5.91 \\
NGC 2976 & ... & 21 & 5 & 0.58 & 7.40 & 610 & 1.70 & 3.90 & 1.90 \\
NGC 3938 & 38 & 30 & 6 & 5.70 & 73.0 & 1750 & 33.0 & 13.5 & 5.41 \\
NGC 4414 & 10 & 4 & 2 & 2.20 & 28.0 & 2740 & 150 & 47.0 & 0.60 \\
NGC 5033 & 35 & 34 & 6 & 23.0 & 294 & 1640 & 111 & 3.20 & 91.9 \\
\tableline
\enddata

\tablenotetext{a}{CO flux in central kiloparsec measured from BIMA SONG}

\tablenotetext{b}{Total gas mass in the central kiloparsec calculated
from the SONG flux. All masses are calculated using M = 1.36 M(H$_2$),
M(H$_2$)=1.1$\times$10$^4$ D$^2$ S$_{CO}$ where D is in Mpc and
S$_{CO}$ is in Jy \kms.  We use a CO-to-H$_2$ coversion factor of
2.8$\times$10$^{20}$ cm$^{-2}$ (K \kms))$^{-1}$, 7\% smaller than
\citet{sakamoto99b}.}
\tablenotetext{c}{Nuclear gas surface density derived by dividing the
mass by $\pi$ (500 pc)$^2$. }
\tablenotetext{d}{Total CO flux from galaxy (disk+nucleus) from the
FCRAO survey \citep{young95}}
\tablenotetext{e}{Total gas mass in the galaxy (disk + nucleus)
calculated from the FCRAO flux}
\tablenotetext{f}{Total gas surface density in the galaxy (disk +
nucleus) calculated from the FCRAO flux}
\tablenotetext{g}{Ratio of the nuclear gas surface density to the disk
gas surface density, defined as f$_{con}$ in \S \ref{intro}}
\tablenotetext{s}{Using galaxy-averaged flux from SONG data instead of the FCRAO survey.}
\tablenotetext{u}{ 2$\sigma$ upper limits}
\end{deluxetable}
\end{landscape}


\clearpage
\end{document}